\documentclass[conference,a4paper]{IEEEtran}

\usepackage{color}
\usepackage{textcomp}
\usepackage[ruled]{algorithm2e}
\usepackage{graphicx}
\usepackage{subfigure}
\usepackage{amsmath,amssymb,amsfonts,mathrsfs}
\usepackage{multirow}
\usepackage{cite}
\usepackage{booktabs}
\usepackage{epstopdf}
\usepackage{xcolor,stfloats}
\usepackage{bm}
\usepackage{subeqnarray}

\DeclareMathAlphabet      {\mathbi}{OT1}{cmr}{bx}{it}
\SetMathAlphabet\mathbi{bold}{OT1}{cmr}{bx}{it}

\newtheorem{definition}{Definition}

\newtheorem{lemma}{Lemma}
\newtheorem{proof}{Proof}

\begin{document}

\title{Sensing-Resistance-Oriented Beamforming for Privacy Protection from ISAC Devices\\
}

\author{\IEEEauthorblockN{Teng Ma\IEEEauthorrefmark{1}, Yue Xiao\IEEEauthorrefmark{1}, Xia Lei\IEEEauthorrefmark{1}, and Ming Xiao\IEEEauthorrefmark{2}}
\IEEEauthorblockA{
\IEEEauthorrefmark{1}National Key Laboratory of Wireless Communications,\\
University of Electronic Science and Technology of China, Chengdu, China\\}
\IEEEauthorblockA{
\IEEEauthorrefmark{2}Department of Information Science and Engineering,\\
Royal Institute of Technology, KTH, Sweden\\}
\IEEEauthorblockA{
Email: \{xiaoyue@uestc.edu.cn\}}
}


\maketitle

\begin{abstract}
With the evolution of integrated sensing and communication (ISAC) technology, a growing number of devices go beyond conventional communication functions with sensing abilities.
Therefore, future networks are divinable to encounter new privacy concerns on sensing, such as the exposure of position information to unintended receivers.
In contrast to traditional privacy preserving schemes aiming to prevent eavesdropping, this contribution conceives a novel beamforming design toward sensing resistance (SR).
Specifically, we expect to guarantee the communication quality while masking the real direction of the SR transmitter during the communication.
To evaluate the SR performance, a metric termed angular-domain peak-to-average ratio (ADPAR) is first defined and analyzed.
Then, we resort to the null-space technique to conceal the real direction, hence to convert the optimization problem to a more tractable form.
Moreover, semidefinite relaxation along with index optimization is further utilized to obtain the optimal beamformer.
Finally, simulation results demonstrate the feasibility of the proposed SR-oriented beamforming design toward privacy protection from ISAC receivers.
\end{abstract}


\IEEEpeerreviewmaketitle
\section{Introduction}
\IEEEPARstart{T}{he} fifth-generation (5G) mobile networks have witnessed the emergence of novel applications with both communication and sensing needs, such as industrial internet of things (IoT), smart home, and internet of vehicles (IoV) \cite{5G}.
Therefore, it is foreseeable that future wireless networks will fuse the services of conventional communication and radio sensing together, while the devices may have both communication and sensing capabilities.
Indeed, some researchers have attempted to combine radar and communication at the base station (BS), by using a dual-function radar and communication (DFRC) BS to serve users while probing echo signals from interested targets simultaneously \cite{DFRC1}.
Then, based on this concept, the technology of integrated sensing and communication (ISAC) arose and hence to attract much attentions \cite{ISAC1}.
Specifically, DFRC BSs have two mainstream architectures: the one is the co-existing design \cite{DFRC2}, where the transmitting wave is the superposition of communication and sensing signals, so the key issue is to handle spectrum sharing and reduce mutual interference; the other is the fusion design \cite{DFRC3}, which seeks to achieve target detection and data transmission via a common waveform.
Whichever, the performance tradeoff between sensing and communication is always the kernel due to the shared use of spectral and infrastructure resources \cite{DFRC4}.

Nowadays, both communication and radio sensing technologies are evolving toward some common directions, including exploiting higher spectral resources such as millimeter and terahertz, deploying massive antenna arrays, developing new electromagnetic materials, and so on \cite{6G1,6G2,6G3}.
Naturally, future networks will integrate more sensing services such as ranging, positioning, speed measurements, imaging, and even environment reconstruction \cite{ISAC2,ISAC3,ISAC4}.
Indeed, ISAC has become a key scene toward the vision of the six-generation (6G) \cite{ISAC5,ISAC6}.
However, the evolution of ISAC with increasing signal processing abilities of devices may introduce new threats on privacy from the perspective of sensing, such as exposure of transmitter's location \cite{Privacy1,Privacy2,Privacy3,Privacy4,PISAC1}, as
radio propagation also carries geometrical information.
Unfortunately, to the best of the authors' knowledge, there are few discussions on privacy in ISAC while current works such as \cite{PISAC2} still concentrated on privacy protection in communication rather than sensing.

Motivated by the above challenges, this contribution conceives a generic sensing resistance (SR)-oriented beamforming (BF) design for privacy protection from ISAC devices.
Specifically, we focus on the data transmission from an SR transmitter to an ISAC receiver while the latter is not allowed to sense the transmitter's real direction.
In other words, the SR transmitter pursues robust communication performance while concealing its real direction information from the ISAC receiver.
In general, the contributions of this paper are summarized as: i) a novel metric of angular-domain peak-to-average ratio (ADPAR) is defined to evaluate the SR performance with further analysis, ii) the closed-form ADPAR bounds are derived through the generalized Rayleigh quotient, iii) we resort to the null-space technique to conceal the real direction, and iv) a semidefinite relaxation (SDR)-based approach along with index optimization is utilized to obtain the optimal beamforming.

The remainder of this paper is organized as follows.
In Section II, the channel model of the conceived SR scheme in the context of ISAC is introduced.
Section III exhibits the proposed SR-BF design in details.
In Section IV, we present simulation results and performance comparisons.
Finally, conclusions are given in Section V.

\emph{Notation:}
In the following, lowercase bold letters and uppercase bold letters represent vectors and matrices, respectively. ${(\cdot)}^{\rm T}$, $(\cdot)^{\rm H}$, $(\cdot)^\ast$ and $(\cdot)^\dagger $stand for the transposition, Hermitian transposition, complex conjugation, and Moore-Penrose generalized inverse, respectively.
$\mathbb C^{m\times n}$ and $\mathbb R^{m\times n}$ stand for the space of $m\times n$ complex and real matrices,
$j=\sqrt{-1}$ is the imaginary number, and $\mathbf I$ denote the identity matrix.
$\rm diag(\mathbf x)$ denotes converting $\mathbf x$ to a diagonal matrix, while $\rm vecd(\cdot)$ represents to extract the diagonal element of a matrix as a column vector.
$\|\cdot\|_F$ is the Frobenius norm while $\mathbf A\succeq \mathbf B$ means $\mathbf A-\mathbf B$ is positive semidefinite.
$\mathcal{CN}(\bm\mu,\bm\Sigma)$ denotes the complex Gaussian distribution with mean $\bm\mu$ and covariance matrix $\mathbf\Sigma$, and $\sim$ stands for ``distributed as''.

\begin{figure}[!t]
\centering
\includegraphics[width=3.5in]{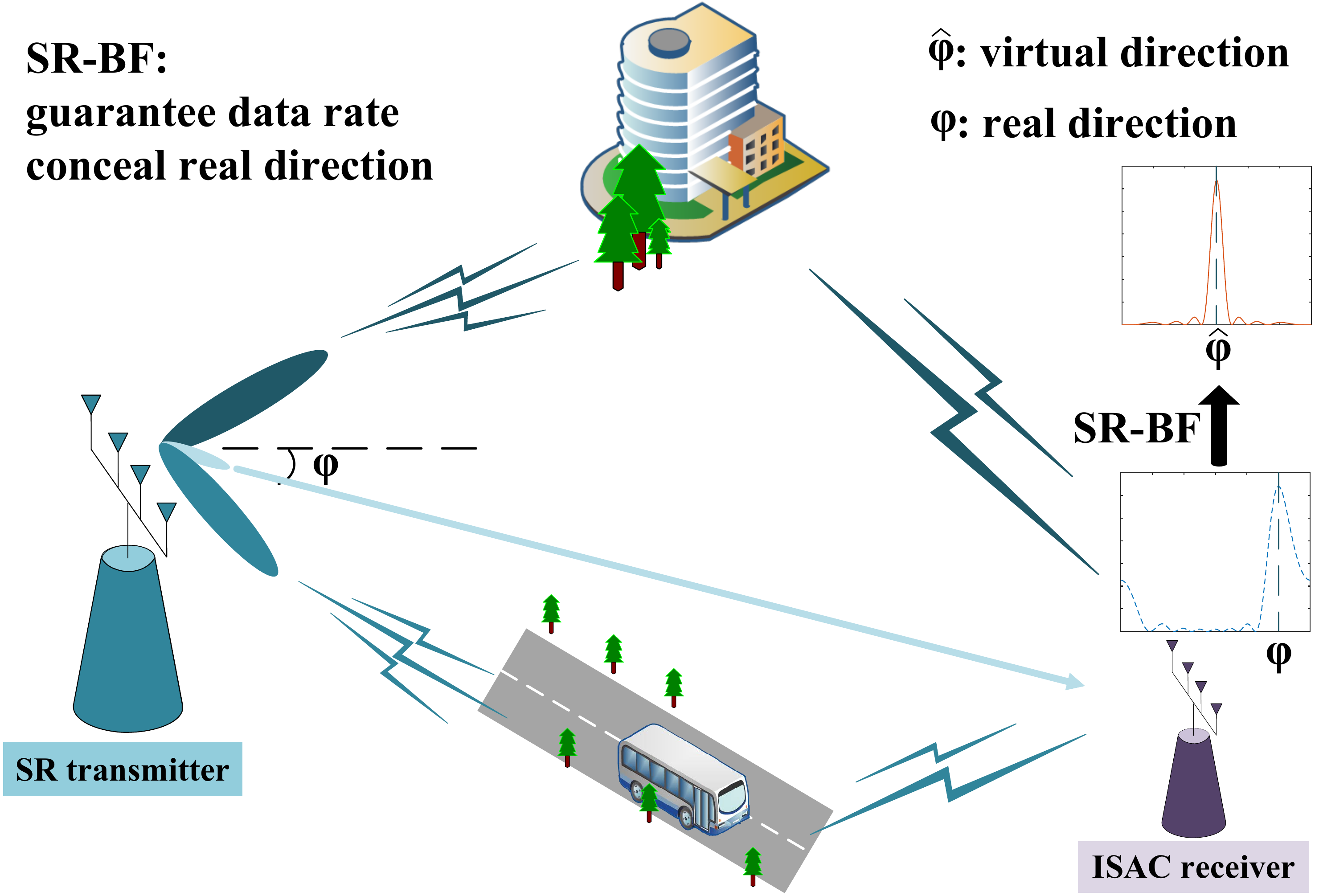}
\vspace{-1em}
\caption{An example of the SR-BF model for privacy protection in ISAC.}
\end{figure}

\section{Signal and Channel Model}
As depicted in Fig. 1, we focus on a two-dimensional (2D) multiple-input multiple-output (MIMO) transmission where an SR transmitter equipped with $N_T$ antennas communicates to an ISAC receiver with $N_R$ antennas.
For the privacy concerns, the SR transmitter expects to guarantee the communication quality while preventing the ISAC receiver from sensing its real direction.
Without loss of generality, we assume block-flat fading in the following, i.e., channel coefficients remain constant within a channel coherence period (CCP) but vary between different CCPs.
Meanwhile, we adopt the widely applicable Rician channel model, i.e., the channel matrix can be expressed as
\begin{equation}
\begin{aligned}
{\bf H}=\alpha\left(\sqrt{\frac{\kappa}{\kappa+1}}
\bar{\bf H}+\sqrt{\frac{1}{\kappa+1}}\tilde{\bf H}\right)\in\mathbb C^{N_R\times N_T},
\end{aligned}
\label{H}
\end{equation}
where $\alpha$ denotes the complex fading coefficient, $\kappa$ is the Rician factor, $\bar{\bf H}$ represents the line-of-sight (LoS) component, and $\tilde{\bf H}$ is the non-LoS (NLoS) component with each entry independently and identically distributed (i.i.d.) to $\mathcal{CN}(0,1)$.
For simplicity, we consider the far-field scenario, i.e., the antenna array response generator can be formulated as
\begin{equation}
{\bf a}_I(\cdot)
=\left[a_{I1}(\cdot),a_{I2}(\cdot),
\dots,a_{IN_I}(\cdot)\right]^{\rm T}, I\in\{T,R\}.
\end{equation}
For example, for a horizontal uniform linear array (ULA), we recall that
\begin{equation}
a_{In}(\cdot)=e^{j\frac{2\pi}{\lambda}(n-1)\Delta \cos(\cdot)},n\in\{1,2,\dots,N_I\},
\end{equation}
where $\lambda$ is the carrier wavelength, while $\Delta$ denotes the element spacing often as $\lambda/2$.
Thus, the LoS component can be determined by the antenna array response, namely
\begin{equation}
\bar{\bf H}={\bf a}_R(\varphi){\bf a}_T^{\rm H}(\varphi).
\label{H_bar}
\end{equation}

To realize SR-BF, the transmitting symbol is precoded by
\begin{equation}
{\bf x}={\bf Ws},
\label{x}
\end{equation}
where ${\bf W}\in\mathbb C^{N_T\times N_S}$ is the precoder that satisfies ${\rm tr}({\bf W}{\bf W}^{\rm H}=P)$ and ${\bf s}\sim\mathcal{CN}(\mathbf 0,\mathbf I)\in\mathbb C^{N_S}$ is the equivalent complex baseband signal, with $P$ and $N_S\leq\min\{N_T,N_R\}$ denoting the transmitting power and the number of data streams, respectively.
Then, the received signal can be expressed as
\begin{equation}
{\bf y}={\bf Hx}+{\bf n}={\bf HWs}+{\bf n},
\label{y}
\end{equation}
where ${\bf n}$ denotes the additive white gaussian noise (AWGN) with each entry i.i.d. to $\mathcal {CN}(0,N_0)$, in which $N_0$ represents the power spectral density (PSD).
According to (\ref{y}), the achievable rate can be formulated by
\begin{equation}
C=\log\det({\bf I}+N_0^{-1}{\bf H}{\bf W}{\bf W}^{\rm H}{\bf H}^{\rm H}).
\label{C}
\end{equation}
On the other hand, to measure the SR performance, we migrate the concept of peak-to-average ratio to the angular domain and give a definition in {\it Definition 1} to describe the ADPAR toward a certain direction.
\definition{The ADPAR in a given direction $\theta$ is expressed as
\begin{equation}
\rho(\theta)\overset{\rm def}=\frac{\mathbf a_R^{\rm H}(\theta)\mathbf R\mathbf a_R(\theta)}{\frac{1}{\pi}\int_{0}^{\pi}{\mathbf a_R^{\rm H}(\theta)\mathbf R\mathbf a_R(\theta)}{\rm d}\theta},
\label{rho}
\end{equation}
where
\begin{equation}
\mathbf R\overset{\rm def}=\mathbb E{\bf y}{\bf y}^{\rm H}={\bf H}{\bf W}{\bf W}^{\rm H}{\bf H}^{\rm H}+N_0{\bf I}
\label{R}
\end{equation}
denotes the spatial covariance matrix of received signals.}

{\it Definition 1} reveals that as $\rho(\theta)$ goes larger, the receiver will have a higher probability to determine $\theta$ as the main direction of incoming signals. Meanwhile, according to the above definition, we have the following result.
\lemma{Denoting $\mathbf A(\theta)=\mathbf a_R(\theta)\mathbf a_R^{\rm H}(\theta)$, $\rho(\theta)$ can be converted to
\begin{equation}
\begin{aligned}
\rho(\theta)=\frac{{\rm tr}\{{\bf W}^{\rm H}[{\bf H}^{\rm H}\mathbf A(\theta){\bf H}+\frac{N_RN_0}{P}\mathbf I]{\bf W}\}}{{\rm tr}[{\bf W}^{\rm H}({\bf H}^{\rm H}\mathbf J{\bf H}+\frac{N_RN_0}{P}\mathbf I){\bf W}]},
\end{aligned}
\end{equation}
where the entry in the $m$-th row and $n$-th column of $\mathbf J$ is given by
\begin{equation}
\begin{aligned}
\left[\mathbf J\right]_{mn}=J_0\left[\frac{2\pi}{\lambda}(m-n)\Delta\right], m,n\in\{1,2,\dots,N_R\}.
\end{aligned}
\end{equation}
\proof{See Appendix A.}
}

This lemma converts the integral and quadratic forms into the trace of matrix products, giving a much simpler expression of ADPAR.
Therefore, we can formulate the following optimization problem to maximize the achievable rate while guaranteeing the privacy, i.e.,
\begin{equation}
\begin{aligned}
(\rm{P1})\ &\max_{\mathbf W}\ C\\
&\ \ {\rm s. t.}\ \ {\rm tr}(\mathbf W\mathbf W^{\rm H})=P,\\
&\qquad\ \ \bar{\bf H}\mathbf W=\mathbf O,\\
&\qquad\ \ \rho(\hat{\varphi})\geq\gamma,\\
\end{aligned}
\label{P1}
\end{equation}
where the second constraint represents to zero-force the LoS signal at the receiver, i.e., preventing it from sensing transmitter's real direction, while the last one guarantees the ADPAR in the virtual direction $(\hat{\varphi})$ at a relatively high level ($\gamma$ is a large positive number). In other words, these two constraints ensure that the true direction of the SR transmitter is concealed by the virtual one.

\section{Sensing-Resistance-Oriented Beamforming}
According to the aforementioned results, it is difficult to solve $(\rm{P1})$ directly due to the non-convex ADPAR constraint.
To conquer it, we first perform some matrix operations to transform $(\rm{P1})$ to a more tractable form.

Note that the second constraint in $(\rm{P1})$ can be converted to
\begin{equation}
\bar{\bf H}\mathbf W=\mathbf O\Leftrightarrow{\bf a}_T^{\rm H}(\varphi)\mathbf W=\mathbf 0,
\end{equation}
which indicates that $\mathbf W$ should belong to the null space of ${\bf a}_T^{\rm H}(\varphi)$.
According to the singular value decomposition (SVD) of ${\bf a}_T^{\rm H}(\varphi)$, i.e.,
\begin{equation}
{\bf a}_T^{\rm H}(\varphi)=
\begin{bmatrix}
\sqrt{N_T}&\mathbf 0\\
\end{bmatrix}
\begin{bmatrix}
\mathbf v_{\max}^{\rm H}\\
\mathbf V_N^{\rm H}
\end{bmatrix},
\end{equation}
we obtain ${\bf a}_T^{\rm H}(\varphi)\mathbf V_N=\mathbf 0$;  in other words, ${\rm Null}[{\bf a}_T^{\rm H}(\varphi)]={\rm Span}(\mathbf V_N)$.
Thus, constructing ${\bf W}=\mathbf V_N{\bf W}'$ with ${\bf W}'\in\mathbb C^{(N_T-1)\times N_S}$, we have ${\bf a}_T^{\rm H}(\varphi)\mathbf W={\bf a}_T^{\rm H}(\varphi)\mathbf V_N\mathbf W'=\mathbf 0$.
Meanwhile, denoting $\mathbf H'=\mathbf H\mathbf V_N$, the achievable rate and ADPAR can be respectively reformulated as
\begin{equation}
C=\log\det({\bf I}+N_0^{-1}{\bf H}'{\bf W}'{\bf W}^{\prime\rm H}{\bf H}^{\prime\rm H}),
\end{equation}
and
\begin{equation}
\begin{aligned}
\rho(\hat{\varphi})
&=\frac{{\rm tr}\{{\bf W}^{\prime\rm H}[{\bf H}^{\prime\rm H}\mathbf A(\hat{\varphi}){\bf H}'+\frac{N_RN_0}{P}\mathbf I]{\bf W}'\}}{{\rm tr}[{\bf W}^{\prime\rm H}({\bf H}^{\prime\rm H}\mathbf J{\bf H}'+\frac{N_RN_0}{P}\mathbf I){\bf W}']}\\
&\overset{\rm def}=\frac{{\rm tr}({\bf W}^{\prime\rm H}\hat{\bf A}'{\bf W}')}{{\rm tr}({\bf W}^{\prime\rm H}{\bf J}'{\bf W}')}.
\end{aligned}
\label{rho_hat}
\end{equation}
Then, $(\rm{P1})$ can be transformed to
\begin{equation}
\begin{aligned}
(\rm{P2})\ &\max_{\mathbf W'}\ C\\
&\ \ {\rm s. t.}\ \ {\rm tr}(\mathbf W'\mathbf W^{\prime\rm H})=P,\\
&\qquad\ \ \rho(\hat{\varphi})\geq\gamma.\\
\end{aligned}
\label{P2}
\end{equation}

To determine an appropriate threshold $\gamma$, we first provide some theoretical analysis on $\rho(\hat{\varphi})$ in the following. Note that $\rho(\hat{\varphi})$ is in the form of generalized Rayleigh quotient, whose boundaries can be given by the generalized eigenvalue decomposition (GEVD).
\definition{
For a given Hermitian matrix ${\bf A}$ and positive definite Hermitian matrix ${\bf B}$, the GEVD of the matrix pari $\{{\bf A},{\bf B}\}$ is expressed as
\begin{equation}
\begin{aligned}
\mathbf T^{\rm H}{\bf A}\mathbf T=\mathbf\Lambda,
\end{aligned}
\end{equation}
and
\begin{equation}
\begin{aligned}
\mathbf T^{\rm H}{\bf B}\mathbf T=\mathbf I,
\end{aligned}
\end{equation}
where $\mathbf T$ is an invertible matrix while $\mathbf\Lambda$ is a non-negative diagonal matrix with
diagonal elements arranged in the descending order.
}

Based on {\it Definition 2}, we have the following result.
\lemma{$\rho(\hat{\varphi})$ is bounded by the smallest and largest generalized eigenvalue of matrix pair $\{\hat{\bf A}',{\bf J}'\}$, i.e.,
\begin{equation}
\begin{aligned}
\lambda_{N_T-1}=\lambda_{\min}\{\hat{\bf A}',{\bf J}'\}
\leq\rho(\hat{\varphi})&\leq\lambda_{\max}\{\hat{\bf A}',{\bf J}'\}
=\lambda_1,
\end{aligned}
\end{equation}
where the first and second equalities hold if
\begin{equation}
\begin{aligned}
{\mathbf W}'{\mathbf W}^{\prime\rm H}=P\frac{\mathbf t_{\min}\{\hat{\bf A}',{\bf J}'\}\mathbf t_{\min}^{\rm H}\{\tilde{\bf A}',{\bf J}'\}}{\mathbf t_{\min}^{\rm H}\{\tilde{\bf A}',{\bf J}'\}\mathbf t_{\min}\{\hat{\bf A}',{\bf J}'\}},
\end{aligned}
\end{equation}
and
\begin{equation}
\begin{aligned}
{\mathbf W}'{\mathbf W}^{\prime\rm H}=P\frac{\mathbf t_{\max}\{\hat{\bf A}',{\bf J}'\}\mathbf t_{\max}^{\rm H}\{\tilde{\bf A}',{\bf J}'\}}{\mathbf t_{\max}^{\rm H}\{\tilde{\bf A}',{\bf J}'\}\mathbf t_{\max}\{\hat{\bf A}',{\bf J}'\}},
\end{aligned}
\end{equation}
where $\mathbf t_{\min}\{\hat{\bf A}',{\bf J}'\}\overset{\rm def}=\mathbf t_{N_T-1}$ and $\mathbf t_{\max}\{\hat{\bf A}',{\bf J}'\}\overset{\rm def}=\mathbf t_1$ denote the generalized eigenvectors corresponding to $\lambda_{N_T-1}$ and $\lambda_1$, respectively.
\proof{See Appendix B.}
}

{\it Lemma 2} reveals that

i. $(\rm{P2})$ has no solution when $\gamma>\lambda_1$;

ii.  when $\gamma=\lambda_1$, the closed-form solution can be obtained by (22);

iii.  when $\gamma\leq\lambda_{N_T-1}$, the ADPAR constraint is inactive so $\mathbf W'_{\rm opt}$ can be obtained by the conventional water-filling algorithm;

iv. when $\gamma\in(\lambda_{N_T-1},\lambda_1)$, $(\rm{P2})$ cannot be solved directly due to the non-convex constraints, which, however, can be handled by the SDR approach.
Specifically, assuming $\mathbf Z=\mathbf W'{\bf W}^{\prime\rm H}$, we arrive at
\begin{equation}
\begin{aligned}
&\left\{\mathbf W'{\bf W}^{\prime\rm H}|\mathbf W'\in\mathbb C^{(N_T-1)\times N_S}\right\}\\
=&\left\{\mathbf Z\in\mathbb C^{(N_T-1)\times(N_T-1)}|\mathbf Z=\mathbf Z^{\rm H}\succeq\mathbf O,{\rm rank}(\mathbf Z)=N_S\right\}.
\end{aligned}
\end{equation}
Therefore, $(\rm{P2})$ can be equivalent to
\begin{equation}
\begin{aligned}
(\rm{P3})\ &\max_{\mathbf Z}\ \log\det\left({\bf I}+N_0^{-1}{\bf H}'\mathbf Z{\bf H}^{\prime\rm H}\right)\\
&\ \ {\rm s. t.}\ \ {\rm tr}\mathbf Z=P,\\
&\qquad\ \ {\rm tr}[(\hat{\bf A}'-\gamma{\bf J}')\mathbf Z]\geq0,\\
&\qquad\ \ \mathbf Z=\mathbf Z^{\rm H}\succeq\mathbf O,{\rm rank}(\mathbf Z)=N_S.\\
\end{aligned}
\label{P3}
\end{equation}
Considering the eigenvalue decomposition (EVD) of $(\hat{\bf A}'-\gamma{\bf J}')$ as
\begin{equation}
\begin{aligned}
(\hat{\bf A}'-\gamma{\bf J}')=\mathbf U'\mathbf\Lambda'\mathbf U^{\prime\rm H},
\end{aligned}
\end{equation}
where $\mathbf U'$ is unitary and $\mathbf\Lambda'={\rm diag}(\lambda'_1,\lambda'_2,\dots,\lambda'_{N_T})$ is diagonal.
Thus, we can let $\mathbf Z=\mathbf U'\mathbf P\mathbf U^{\prime\rm H}$, where $\mathbf P={\rm diag}(p_{1},p_{2},\dots,p_{N_T})$ is non-negative and and diagonal with ${\rm rank}(\mathbf P)=N_S$. Denoting the indices of non-negative entries of $\mathbf P$ as $\mathcal I$, $(\rm{P3})$ can be decompose into two subproblems as
\begin{equation}
\begin{aligned}
(\rm{P3-1})\ &\max_{\mathcal I}\ \log\det\left[{\bf I}+N_0^{-1}{\bf H}'\mathbf U'_{:,\mathcal I}\mathbf P'(\mathcal I){\bf U}^{\prime\rm H}_{:,\mathcal I}{\bf H}^{\prime\rm H}\right],\\
&\ \ {\rm s. t.}\ \ \mathcal I\subseteq\{1,2\dots,N_T-1\},\\
&\qquad\ \ |\mathcal I|=N_S,\\
\end{aligned}
\label{P3-1}
\end{equation}
where $\mathbf P'(\mathcal I)={\rm diag}(p_{\mathcal I_1},p_{\mathcal I_2},\dots,p_{\mathcal I_{N_S}})$ is the submatrix formulated by the non-negative entries of $\mathbf P$, which can be obtained by solving the following concave problem:
\begin{equation}
\begin{aligned}
(\rm{P3-2})\ &\max_{\{p_i>0\}}\ \log\det\left({\bf I}+N_0^{-1}{\bf H}'\mathbf U'_{:,\mathcal I}\mathbf P'{\bf U}^{\prime\rm H}_{:,\mathcal I}{\bf H}^{\prime\rm H}\right)\\
&\ \ {\rm s. t.}\ \ \sum_{i=1}^{N_S}p_{\mathcal I_i}=P,\\
&\qquad\ \ \sum_{i=1}^{N_S}p_{\mathcal I_i}\lambda'_{i}\geq0.\\
\end{aligned}
\label{P3-2}
\end{equation}
Consequently, the optimal beamformer can be constructed by
\begin{equation}
\mathbf W'_{\rm opt}=\mathbf U'_{:,\mathcal I_{\rm opt}}\sqrt{{\mathbf P}'(\mathcal I_{\rm opt})}.
\end{equation}
\section{Simulation Results}
\addtolength{\topmargin}{0.01cm}

\begin{figure}[!t]
\centering
\subfigure[Maximal ADPARs.]{
\includegraphics[width=1.625in]{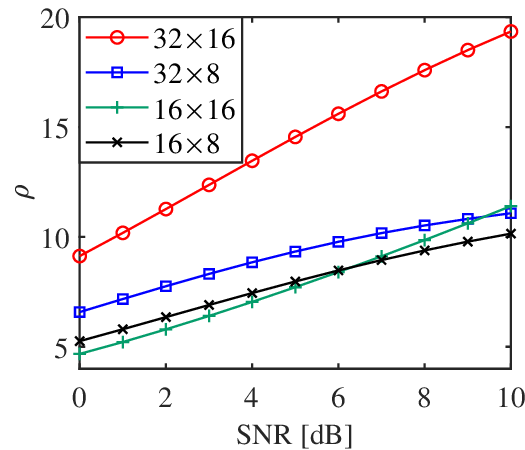}
}
\subfigure[Corresponding maximal achievable rates.]{
\includegraphics[width=1.625in]{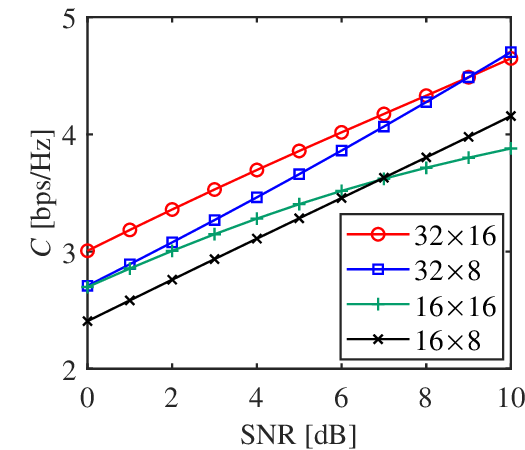}
}
\caption{Maximal ADPARs and corresponding maximal achievable rates with respect to SNR under $N_T=32,16$, and $N_R=16,8$.}
\end{figure}

In the following, we present several numerical examples to demonstrate the feasibility of the proposed SR-oriented beamforming design.
Specifically, both the transmitter and receiver are assumed to equip a horizontal ULA with half-wavelength spacing between adjacent antennas while the transmit power is 1 W.
Meanwhile, we define SNR as $\frac{P|\alpha|^2}{N_0}$ while setting $\varphi=60^\circ$ and $\hat{\varphi}=90^\circ$ and $\kappa=0$ dB.

Fig. 2 provides the results of maximal ADPARs and corresponding maximal achievable rates with respect to SNR under $N_T=32,16$, and $N_R=16,8$. It is shown that both maximal ADPAR and corresponding maximal achievable rate increase as SNR goes up, but with a decreasing growth rate. Meanwhile, it is also exhibited that a higher $N_R$ indicates higher maximal ADPAR at high SNRs, but the phenomenon is reversed at low SNRs. For example, the maximal ADPAR under $16\times16$ is larger than that under $16\times8$ when SNR$>$6 dB, but becomes smaller when SNR$<$6 dB. Interestingly, the influence of $N_R$ on corresponding maximal achievable rate is opposite.
Moreover, we also find that the impact of $N_T$ is simpler; that is, a larger $N_T$ results in larger maximal ADPAR and corresponding maximal achievable rate.

On the other hand, Fig. 3 illustrates the influence of ADPAR threshold $\gamma$ on the maximal achievable rate and beampatterns, compared with the conventional benchmarks (maximizing the achievable rate by the water-filling approach). Specifically, these results are plotted under $N_T=16$, $N_R=8$, $N_S=4$, and $\gamma=0,5,\rho_{\max}$. As expected, the SBR schemes have lower rates than the conventional benchmark, since the LoS channel component is zero-forced to hide the angular information. Meanwhile, it is also depicted that a larger $\gamma$ indicates a smaller rate but a sharper peak in the direction of $\hat{\varphi}$, since more power will be allocated to the desired direction of for angle cheating.

\begin{figure}[!t]
\centering
\subfigure[Maximal achievable rates.]{
\includegraphics[width=1.625in]{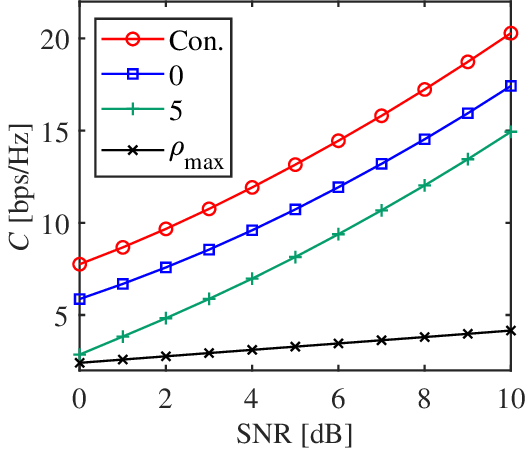}
}
\subfigure[Beampatterns under SNR = 10 dB.]{
\includegraphics[width=1.625in]{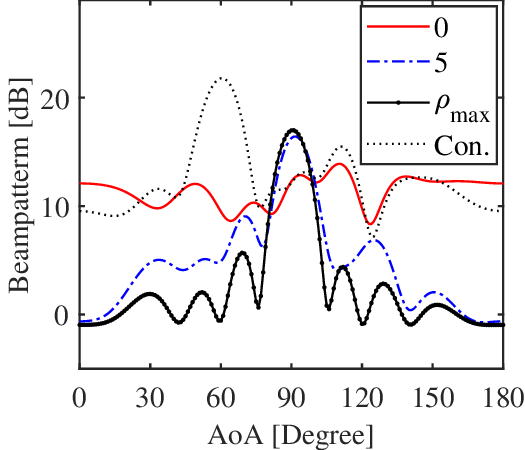}
}
\caption{Maximal achievable rates and beampatterns under $N_T=16$, $N_R=8$, $N_S=4$, and $\gamma=0,5,\rho_{\max}$, compared with conventional benchmarks.}
\end{figure}

\begin{figure}[!t]
\centering
\subfigure[Maximal achievable rates under $N_S=1,2,4,6,8$]{
\includegraphics[width=1.625in]{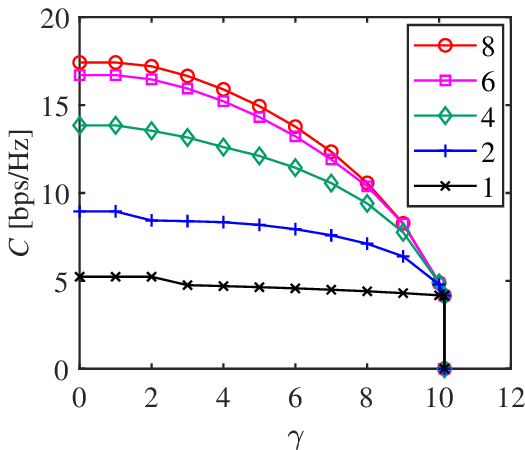}
}
\subfigure[Beampatterns under $N_S=1,4,8$ and $\gamma=5$.]{
\includegraphics[width=1.625in]{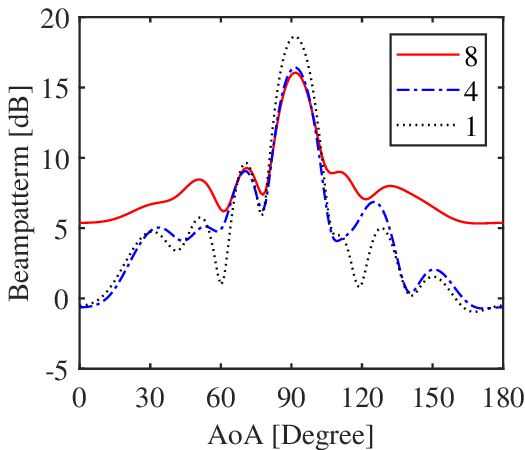}
}
\caption{Maximal achievable rates and beampatterns under $N_T=16$, $N_R=8$, and SNR $=10$ dB.}
\end{figure}

By contrast, Fig. 4 exhibits the simulation results under different $N_S$ given $\gamma=5$ and SNR = 10 dB to show the influence of the number of data streams. Evidently, the achievable rate increases as $N_S$ goes up under the same parameter settings, but with a decreasing growth rate. For example, at $\gamma=4$, the rate increases by about 4 bps/Hz from $N_S=2$ to 4 and increases by about 3 bps/Hz from $N_S=4$ to 8. Meanwhile, it is also shown that decreasing $N_S$ leads to lower sidelobe level, indicating that enlarging the rate and achieving better angel cheating performance is a tradeoff.

\section{Conclusions}
In this work, we investigated an SR-oriented beamforming design for privacy protection from ISAC devices, where the SR transmitter tends to improve
the achievable rate while preventing the ISAC receiver from sensing its real direction during communication.
Specifically, we firstly defined a metric termed ADPAR to evaluate the SR performance and adopted the generalized Rayleigh quotient to determine its boundaries.
Then, the optimization problem was formulated under ADPAR constraints, where the null-space technique was adopted to conceal the real direction of the SR
transmitter. Then, we resorted to SDR and index optimization approaches to obtain the optimal beamforming.
Simulation results demonstrated the feasibility of the proposed SR-oriented beamforming design toward
privacy protection from ISAC receivers.

\section{Acknowledgement}
This work was supported in part by the National Key R\&D Program of China under Grant 2023YFE0115100, National Science Foundation of China under Grant number 62071111, and the National Key Laboratory of Wireless Communications Foundation under Grant IFN20230104.

\begin{appendices}
\section{}
\setcounter{equation}{0}
\renewcommand{\theequation}{A\arabic{equation}}
According to ${\rm tr}(\mathbf A\mathbf B)={\rm tr}(\mathbf B\mathbf A)$, (\ref{rho}) can be converted to
\begin{equation}
\begin{aligned}
\rho(\theta)&=\frac{\mathbf a_R^{\rm H}(\theta)\mathbf R\mathbf a_R(\theta)}{\frac{1}{\pi}\int_{0}^{\pi}{\mathbf a_R^{\rm H}(\theta)\mathbf R\mathbf a_R(\theta)}{\rm d}\theta}
=\frac{{\rm tr}[\mathbf R\mathbf A(\theta)]}{\frac{1}{\pi}\int_{0}^{\pi}{\rm tr}[{\mathbf R\mathbf A(\theta)}]{\rm d}\theta}\\
&=\frac{{\rm tr}[\mathbf R\mathbf A(\theta)]}{{\rm tr}[\mathbf R\int_{0}^{\pi}\frac{1}{\pi}{\mathbf A(\theta)}{\rm d}\theta]}.
\end{aligned}
\end{equation}
For the entry in the $m$-th row and $n$-th column of $\mathbf A(\theta)$, we have
\begin{equation}
\begin{aligned}
\int_{0}^{\pi}\frac{1}{\pi}\left[\mathbf A(\theta)\right]_{m,n}{\rm d}\theta
&=\int_{0}^{\pi}\frac{1}{\pi}e^{j\frac{2\pi}{\lambda}(m-n)\Delta}{\rm d}\theta\\
&=J_0\left[\frac{2\pi}{\lambda}(m-n)\Delta\right]\overset{\rm def}=[\mathbf J]_{m,n},
\end{aligned}
\end{equation}
where $m,n\in\{1,2,\dots,N_R\}$, which further yields
\begin{equation}
\begin{aligned}
\rho(\theta)&=\frac{{\rm tr}[\mathbf R\mathbf A(\theta)]}{{\rm tr}(\mathbf R\mathbf J)}=\frac{{\rm tr}\{{\bf W}^{\rm H}[{\bf H}^{\rm H}\mathbf A(\theta){\bf H}+\frac{N_RN_0}{P}\mathbf I]{\bf W}\}}{{\rm tr}[{\bf W}^{\rm H}({\bf H}^{\rm H}\mathbf J{\bf H}+\frac{N_RN_0}{P}\mathbf I){\bf W}]}.
\end{aligned}
\end{equation}
This completes the proof.
\section{}
\setcounter{equation}{0}
\renewcommand{\theequation}{B\arabic{equation}}
According to (\ref{rho_hat}), it is apparent that both $\hat{\bf A}'$ and ${\bf J}'$ are positive definite Hermitian, indicating that the GEVD exists.
Denote ${\bf W}'=[\mathbf w'_1,\mathbf w'_1,\dots,\mathbf w'_{N_T-1}]$, we have
\begin{equation}
\begin{aligned}
\rho(\hat{\varphi})
=\frac{{\rm tr}({\bf W}^{\prime\rm H}\hat{\bf A}'{\bf W}')}{{\rm tr}({\bf W}^{\prime\rm H}{\bf J}'{\bf W}')}
=\frac{\sum_{i=1}^{N_T-1}{\bf w}_i^{\prime\rm H}\hat{\bf A}'{\bf w}_i'}{\sum_{i=1}^{N_T-1}{\bf w}_i^{\prime\rm H}{\bf J}'{\bf w}_i'}.
\end{aligned}
\end{equation}
According to the generalized Rayleigh-Ritz quotient theorem \cite{Toolbook1}, we obtain
\begin{equation}
\begin{aligned}
\lambda_{\min}\{\hat{\bf A}',{\bf J}'\}\leq\frac{{\bf w}_i^{\prime\rm H}\hat{\bf A}'{\bf w}_i'}{{\bf w}_i^{\prime\rm H}{\bf J}'{\bf w}_i'}\leq\lambda_{\max}\{\hat{\bf A}',{\bf J}'\},
\end{aligned}
\end{equation}
where the first and second equalities hold if ${\bf w}_i'=x_i\mathbf t_{\min}\{\hat{\bf A}',{\bf J}'\}$ and ${\bf w}_i'=x_i\mathbf t_{\max}\{\hat{\bf A}',{\bf J}'\}$ with $x_i\neq0$. Thus, we have
\begin{equation}
\begin{aligned}
\rho(\hat{\varphi})\geq\frac{\lambda_{\min}(\hat{\bf A}',{\bf J}')\sum_{i=1}^{N_T-1}{\bf w}_i^{\prime\rm H}{\bf J}'{\bf w}_i'}{\sum_{i=1}^{N_T-1}{\bf w}_i^{\prime\rm H}{\bf J}'{\bf w}_i'}=\lambda_{\min}(\hat{\bf A}',{\bf J}'),
\end{aligned}
\end{equation}
and
\begin{equation}
\begin{aligned}
\rho(\hat{\varphi})\leq\frac{\lambda_{\max}(\hat{\bf A}',{\bf J}')\sum_{i=1}^{N_T-1}{\bf w}_i^{\prime\rm H}{\bf J}'{\bf w}_i'}{\sum_{i=1}^{N_T-1}{\bf w}_i^{\prime\rm H}{\bf J}'{\bf w}_i'}
=\lambda_{\max}(\hat{\bf A}',{\bf J}').
\end{aligned}
\end{equation}
This completes the proof.
\end{appendices}

\end{document}